\documentstyle[12pt,fleqn]{article}
\def\soc{{\rm C}_{60}}
\def\rug{{\rm C}_{70}}
\def\c76{{\rm C}_{76}}
\def\beeq{\begin{equation}}
\def\eneq{\end{equation}}
\def\beeqa{\begin{eqnarray}}
\def\eneqa{\end{eqnarray}}

\addtocounter{section}{-1}
\begin{document}

\begin{center}

\vspace{2in}

{\Large {\bf Optical Absorption in Higher Fullerenes:\\
Effects of Symmetry Reduction}}\\

\vspace{1cm}

{\rm Kikuo H{\sc arigaya}\footnote[1]{E-mail address: harigaya@etl.go.jp.}
}\\
\mbox{} \\
{\sl Fundamental Physics Section, Electrotechnical Laboratory,\\
Umezono 1-1-4, Tsukuba 305}\\

\vspace{1cm}

(Received March 18, 1994)

\vspace{1cm}
\end{center}

\noindent
{\Large {\bf Abstract}}\\ %
Optical absorption spectra of higher fullerenes ($\rug$ and $\c76$)
as well as of $\soc$ are calculated by a tight binding model with a
long range Coulomb interaction.  A reasonable parameter set gives
calculated spectra which are in overall agreement with the
experiments of of $\soc$ and $\rug$ in solutions.  The variations
of the spectral shape are discussed relating with the symmetry
reduction going from $\soc$ and $\rug$ to $\c76$: the optical gap
decreases and the spectra have more small structures in the
dependences on the excitation energy.

\mbox{}


\pagebreak


Recently, the fullerene family C$_N$ with hollow cage structures
have been intensively investigated.  Various optical experiments have been
performed, and interesting properties due to $\pi$ electrons
delocalized on molecule surfaces have been revealed.  They include
the optical absorption spectra of $\soc$ and $\rug$,$^{1,2)}$ and the
large optical nonlinearity of $\soc$.$^{3,4)}$  The absorption
spectra of higher fullerenes ($\c76$, C$_{78}$, etc.) have been
obtained also.$^{5,6)}$  In order to analyze the optical
properties, we have studied the linear absorption and the third
harmonic generation of $\soc$ by using a tight binding model$^{7)}$
and a model with a long range Coulomb interaction.$^{8)}$

The purpose of this paper is to extend the calculation of $\soc$$^{8)}$
up to one of higher fullerenes, $\c76$.  The main purpose is to look at
how the optical absorption changes as the symmetry of the molecule
reduces going from $\soc$ and $\rug$ to $\c76$.  There have been two possible
isomers for $\c76$ which satisfy the isolated pentagon rule.$^{9)}$
They have $D_2$ [Fig. 1(a)] and $T_d$ symmetries [Fig. 1(b)].
Owing to the NMR experiment,$^{5)}$ the molecular structure
has been identified.  There is only the molecule with the $D_2$ symmetry.
We, however, calculate the optical absorption for both of them
in order to look at symmetry reduction effects extensively.

We use the following hamiltonian:
\beeq
H = H_0 + H_{\rm bond} + H_{\rm int}.
\eneq
The first term of eq. (1) is the tight binding model:
\beeq
H_0 = - t \sum_{\langle i,j \rangle, \sigma}
(c_{i,\sigma}^\dagger c_{j,\sigma} + {\rm h.c.}),
\eneq
where $t$ is the hopping integral and $c_{i,\sigma}$ is an annihilation
operator of a $\pi$-electron with spin $\sigma$.  If $t$ depends on
the bond length, the results do not change so strongly, because main
contributions come from the strong Coulomb potential.  Effects of zero
point vibrations and thermal fluctuation of the lattice are described
by the bond disorder model which is the second term of eq. (1):
\beeq
H_{\rm bond} = \sum_{\langle i,j \rangle, \sigma} \delta t_{i,j}
(c_{i,\sigma}^\dagger c_{j,\sigma} + {\rm h.c.}).
\eneq
Here, $\delta t_{i,j}$ is the Gaussian disorder potential at the
bond $\langle i,j \rangle$.   We can estimate the strength of the
disorder (standard deviation) $t_{\rm s}$ from the results by the
extended Su-Schrieffer-Heeger model.$^{10)}$  The value would
be $t_{\rm s} \sim 0.05 - 0.1 t$.  This is of the similar magnitude
as in the fullerene tubules and conjugated polymers.  We shall
treat interactions among $\pi$-electrons by the following model:
\beeqa
H_{\rm int} &=& U \sum_i
(c_{i,\uparrow}^\dagger c_{i,\uparrow} - \frac{1}{2})
(c_{i,\downarrow}^\dagger c_{i,\downarrow} - \frac{1}{2})\\ \nonumber
&+& \sum_{i \neq j} W(r_{i,j})
(\sum_\sigma c_{i,\sigma}^\dagger c_{i,\sigma} - 1)
(\sum_\tau c_{j,\tau}^\dagger c_{j,\tau} - 1),
\eneqa
where $r_{i,j}$ is the distance between the $i$th and $j$th sites and
\beeq
W(r) = \frac{1}{\sqrt{(1/U)^2 + (r/r_0 V)^2}}
\eneq
is the Ohno potential.  The quantity $U$ is the strength of
the onsite interaction, $V$ means the strength of the long range
Coulomb interaction, and $r_0$ is the average bond length.

The model is treated by the Hartree-Fock approximation and the
single excitation configuration interaction method, as we used
in the previous papers.$^{8,11)}$   In ref. 11, we
have varied the parameters of the Coulomb interaction and the disorder
potential, and have searched for the data which reproduce overall
features of experiments of $\soc$ and $\rug$ in solutions.  We have
found that the common parameters, $U = 4t$, $V = 2t$, $t_{\rm s} = 0.09t$,
and $t=1.8$eV, are reasonable.  We use the same parameter
set for higher fullerenes, too.  For $\rug$, the anisotropic
spectra with respect to the orientation of the molecule against
the electric field of light, as reported in the free electron
model (H\"{u}ckel theory),$^{12)}$ have been averaged by rotating
the molecule.  The averaged data can be compared with experiments
of molecules in solutions.$^{1,2)}$   We use the same procedure for
$\c76$.  Samples of the orientation of the molecule and the
disorder potential are changed 100 times.  This is enough to
obtain smooth numerical data.

First, we discuss the calculated optical absorption of $\soc$ and $\rug$.
The $\soc$ has the high $I_h$ symmetry, and $\rug$ has the lower
$D_{5h}$ symmetry.  Figure 2(a) shows the spectrum of $\soc$,
and Fig. 2(b) displays that of $\rug$.  Experimental data are taken
from refs. 1 and 2, and are shown by thin lines.  Here, we concentrate upon
effects of the symmetry reduction from $I_h$ to $D_{5h}$.  There
are three main features in the $\soc$ absorption.  They are around
the energies, 3.5eV, 4.7eV, and 5.6eV.  When we turn to $\rug$,
we could say that several small peaks in the energy interval
from 1.7eV to 3.6eV come from the 3.5eV feature of $\soc$ after
splitting.  Both the 4.7eV and 5.6eV features of $\soc$ join
into the large feature which is present in the energy region
larger than 3.6eV.  The optical gap decreases from 3.1eV
($\soc$) to 1.7eV ($\rug$).  These changes would be due to the
symmetry reduction from the $\soc$ soccerball to the $\rug$ rugbyball.

Next, we show optical spectra of $\c76$.  Figures 3 (a) and (b)
show the calculated results for the $D_2$ and $T_d$ symmetries,
respectively.  The average by the rotation has been performed.
The optical gap is about 1.2eV in Fig. 3(a) and about 0.7eV
in Fig. 3(b).  These values are smaller than the optical gaps
of $\soc$ and $\rug$.  In Fig. 3(a), there are two broad large features
around 2.7eV and 4.5eV with fine structures.  We could say that several
small peaks around 2.7eV and 4.7eV in the $\rug$ data become
fragmented to form the two wide features in the $\c76$ data.
There is still a dip near 3.2eV as in $\rug$.  The position of
the dip does not change so much.  In Fig. 3(b), the spectral shape
becomes broad totally and many small structures develop as well.
These variations could be regarded as owing to the further reduction
of the symmetry of the higher fullerene $\c76$.

The experiments$^{5,6)}$ show that the optical gap decreases
in higher fullerenes (C$_{76}$, C$_{78}$, C$_{82}$, and so on),
and thus this property is consistent with the present calculation.
However, many small structures in the absorption of $\c76$$^{5)}$
which is shown in Fig. 3(a) are not so apparent.  Therefore it seems
difficult to compare with calculations in detail.  The agreement
with experiments is worse for higher fullerenes.  This fact
indicates that the model with only $\pi$ electrons is so simple
for larger fullerenes.  The mixing between $\pi$ and $\sigma$ orbitals
is different for each carbon atom due to the low symmetries of $\rug$
and $\c76$.  Inclusion of the $\sigma$ orbitals would be desirable
for more detailed comparison with experiments.

There is a common trend in $\soc$, $\rug$, and $D_2$-$\c76$:
the deviation of the theory from experiments is larger in the
excitation energy region 5.0-6.0eV.  Excitations among $\sigma$
orbitals would be mixed in this energy region.  This effects
could be taken into account by using models with $\pi$ and
$\sigma$ electrons.

We could note the following point, even though the present model turned
out to be so simple for higher fullerenes.  As is evident for the two
possible isomers of $\c76$, the calculated spectral shapes are different
each other.  Therefore, optical measurements are good methods for
identification of various isomers of more higher fullerenes
(C$_{78}$, C$_{82}$, etc.), for which the number of isomers
rapidly increases.

To summarize, we have considered optical spectra of the higher
fullerenes, $\rug$ and $\c76$, as well as the buckyball $\soc$.
A reasonable parameter set gives calculated spectra which are in overall
agreement with the experiments of of $\soc$ and $\rug$ in solutions.
We have mainly looked at how the optical absorption changes as
the symmetry of the molecule reduces from $\soc$ and $\rug$ to $\c76$.
We have pointed out that the optical gap decreases and the spectra
have more small structures in the energy dependences as the number
of carbons increases.

\mbox{}

\noindent
{\Large {\bf Acknowledgements}}\\
The author acknowledges helpful discussion and correspondence with
Dr. Shuji Abe and Dr. Mitsutaka Fujita.  He also thanks
Mr. Mitsuho Yoshida for providing him with molecular coordinates
of $\rug$ and $\c76$ which are generated by the program$^{13)}$
based on the projection method on the triangular lattice$^{13)}$
or on the honeycomb lattice.$^{14)}$


\mbox{}

\begin{flushleft}
{\Large {\bf References}}
\end{flushleft}

\noindent
1) J. P. Hare, H. W. Kroto, and R. Taylor: Chem. Phys. Lett.
{\bf 177} (1991) 394.\\
2) S. L. Ren, Y. Wang, A. M. Rao, E. McRae, J. M. Holden, T. Hager,
KaiAn Wang, W. T. Lee, H. F. Ni, J. Selegue, and P. C. Eklund:
Appl. Phys. Lett. {\bf 59} (1991) 2678.\\
3) J. S. Meth, H. Vanherzeele, and Y. Wang: Chem. Phys. Lett.
{\bf 197} (1992) 26.\\
4) Z. H. Kafafi, J. R. Lindle, R. G. S. Pong, F. J. Bartoli,
L. J. Lingg, and J. Milliken: Chem. Phys. Lett. {\bf 188} (1992) 492.\\
5) R. Ettl, I. Chao, F. Diederich, and R. L. Whetten:
Nature {\bf 353} (1991) 149.\\
6) K. Kikuchi, N. Nakahara, T. Wakabayashi, M. Honda, H. Matsumiya,
T. Moriwaki, S. Suzuki, H. Shiromaru, K. Saito, K. Yamauchi,
I. Ikemoto, and Y. Achiba: Chem. Phys. Lett. {\bf 188} (1992) 177.\\
7) K. Harigaya and S. Abe: Jpn. J. Appl. Phys. {\bf 31} (1992) L887.\\
8) K. Harigaya and S. Abe: J. Lumin. (1994) to be published.\\
9) D. E. Manolopoulos: J. Chem. Soc. Faraday Trans. {\bf 87} (1991) 2861.\\
10) B. Friedman and K. Harigaya: Phys. Rev. B {\bf 47} (1993) 3975;
K. Harigaya: Phys. Rev. B {\bf 48} (1993) 2765.\\
11) K. Harigaya and S. Abe: Phys. Rev. B (1994) to be published.\\
12) J. Shumway and S. Satpathy: Chem. Phys. Lett. {\bf 211} (1993) 595.\\
13) M. Yoshida and E. \={O}sawa: {\sl The Proceedings of the 3rd IUMRS
International Conference on Advanced Materials, 1993}, to be published;
M. Yoshida and E. \={O}sawa: {\sl The Japan Chemistry Program Exchange},
Program No. 74.\\
14) M. Fujita, R. Saito, G. Dresselhaus, and M. S. Dresselhaus:
Phys. Rev. B {\bf 45} (1992) 13834.\\

\pagebreak

\begin{flushleft}
{\bf Figure captions}
\end{flushleft}

\noindent
Fig. 1.  Molecular structures for $\c76$ with (a) $D_2$ and
(b) $T_d$ symmetries.

\mbox{}

\noindent
Fig. 2.  Optical absorption spectra for (a) $\soc$ and (b) $\rug$,
shown in the arbitrary units.  The abscissa is scaled by $t$.
The parameters in the Ohno potential are $U = 4t$ and $V = 2t$
($t=1.8$eV).   Experimental data are shown by thin lines.  They
are taken from ref. 2  for (a), and from ref. 1 for (b).

\mbox{}

\noindent
Fig. 3.  Optical absorption spectra for $\c76$ with (a) $D_2$ and
(b) $T_d$ symmetries, shown in the arbitrary units.
The abscissa is scaled by $t$.  Parameters in the Ohno potential
are $U = 4t$ and $V = 2t$ ($t = 1.8$eV).  Experimental data of
$D_2$-$\c76$ are shown by the thin line.  They are taken from ref. 5.

\end{document}